\documentclass[review]{elsarticle}
\usepackage{lineno,hyperref}
\usepackage{comment}
\modulolinenumbers[5]
\journal{Journal of \LaTeX\ Templates}

\biboptions{sort&compress}
\usepackage{graphics}
\usepackage{epsfig}
\usepackage{color}

\usepackage{graphicx}

\begin{document}

\begin{frontmatter}

\title{Characterization of the SIDDHARTA-2 luminosity monitor}

\author[IPJ]{M.~Skurzok\fnref{INFN}}  
\author[INFN]{A.~Scordo} 
\author[IPJ]{S.~Niedzwiecki} 
\author[Milano]{A.~Baniahmad} 
\author[INFN]{M.~Bazzi}
\author[Zagreb]{D.~Bosnar}
\author[Romania]{M.~Bragadireanu}
\author[Milano]{M.~Carminati}
\author[Austria]{M.~Cargnelli}
\author[INFN]{A.~Clozza}
\author[INFN]{C.~Curceanu}
\author[INFN]{L.~De Paolis}
\author[INFN,CFermi]{R.~Del Grande}
\author[Munich]{L.~Fabbietti}
\author[Milano]{C.~Fiorini}
\author[INFN]{C.~Guaraldo}
\author[INFN]{M.~Iliescu}
\author[Riken]{M.~Iwasaki}
\author[INFN]{P.~Levi Sandri}
\author[Austria]{J.~Marton}
\author[INFN]{M.~Miliucci}
\author[IPJ]{P.~Moskal}
\author[CFermi,INFN]{K.~Piscicchia}
\author[INFN]{F.~Sgaramella}
\author[INFN,Austria]{H.~Shi}
\author[IPJ]{M.~Silarski}
\author[INFN,Romania]{D.~L.~Sirghi}
\author[INFN,Romania]{F.~Sirghi}
\author[INFN]{A.~Spallone}
\author[Austria]{M.~T\"uchler}
\author[INFN,Munich]{O.~Vazquez Doce}
\author[Austria]{J.~Zmeskal}

\address[INFN]{INFN, Laboratori Nazionali di Frascati, Via E. Fermi, 54, 
 00044 Frascati (Roma), Italy} 
 
 \address[IPJ]{Institute of Physics, Jagiellonian University, prof.\ Stanis{\l}awa {\L}ojasiewicza~11, 30-348 Krak\'{o}w, Poland}
 
\address[Milano]{Politecnico di Milano, Dipartimento di Elettronica, Informazione e Bioingegneria and INFN Sezione di Milano, Milano, Italy}

\address[Zagreb]{Department of Physics, Faculty of Science, University of Zagreb, Zagreb, Croatia}

\address[Romania]{Horia Hulubei National Institute of Physics and Nuclear Engineering (IFIN-HH) Măgurele, Romania}

\address[Austria]{Stefan-Meyer-Institut für Subatomare Physik, Vienna, Austria}

\address[CFermi]{CENTRO FERMI — Museo Storico della Fisica}

\address[Munich]{Excellence Cluster Universe, Technische Universiät München Garching, Germany}

\address[Riken]{RIKEN, Tokyo, Japan}

\begin{abstract}
A luminosity monitor, based on plastic scintillator detectors, has been developed for the SIDDHARTA-2 experiment aiming to perform high precision measurements of kaonic atoms and was installed in 2020 on the DA$\Phi$NE $e^+e^-$ collider at LNF (Laboratori Nazionali di Frascati, INFN). The main goal of this system is to provide the~instantaneous and integrated luminosity of the DA$\Phi$NE facility by measuring the rate of K$^{+}$K$^{-}$ correlated pairs emitted by the $\phi$ meson decay. This task requires an accurate timing of the DAQ signals, as well as timing resolution below 1ns, in order to disentangle the K$^{\pm}$ signals from the background minimum ionizing particles (MIPs) produced during the $e^{+}$ $e^{-}$ collisions at DA$\Phi$NE. In this paper the luminosity monitor concept as well as its laboratory characterization and the first results inside DA$\Phi$NE are presented.
\end{abstract}


\end{frontmatter}

\section{Introduction}

The DA$\Phi$NE (Double Annular $\phi$ Factory for Nice Experiments) accelerator complex~\cite{dafne1,dafne2} at the Frascati National Laboratories of INFN (LNF-INFN), has been designed for low-energy kaon experiments.~DA$\Phi$NE provides a unique quality kaon beam generated by the decay of the $\phi$(1020) mesons, formed almost at rest via $e^{+}e^{-}$ collisions, achieving peak luminosity values in the order of $\sim10^{32}$~cm$^{-2}$s$^{-1}$~\cite{GalloCP2006}. The $\phi \rightarrow$ K$^{+}$K$^{-}$ decays (BR=48.9\%~\cite{pdg}) deliver monochromatic ($\Delta p_{K}/p_{K}$ $<$ 0.1\%) charged kaons with low-momentum ($p_{K}\sim 127$~MeV/c) providing an ideal beam for kaonic atoms measurements. 

These exotic atomic systems, where an electron is replaced by a negatively charged kaon, are currently one of the hottest topics in the field of nuclear and hadronic strangeness physics, both from the experimental and theoretical points of view~\cite{CurceanuRMP20219,DaviesPLB1979,IzyckiZPA1980,BeerPRL2005,IwasakiPRL1997,SchevchenkoNPA2012,FriedmanNPA2017,FriedmanNPA2012,FriedmanHI2012,MizutaniPRC2013,WeisePRC2017,HoshinoPRC2017,DoringPLB2011,GalJMPA,KamalovNPA2001,BazziPLB2011a,BazziPLB2011b,siddh3,siddh4}.~The investigation of kaonic atoms, in particular of the lightest ones, allows to obtain precise information on the antikaons-nucleons/nuclei interaction. In particular, from their radiative cascade transitions measurements isospin-dependent antikaon-nucleon scattering lengths can be extracted, providing fundamental input for understanding the low energy QCD in the strangeness sector, which has an important impact in nuclear and particle physics as well as in astrophysics (equations of state (EoS) of neutron stars)~\cite{KaplanPLB1987,ScordoAIP2016}.~These isospin-dependent scattering lengths can be extracted from the precise measurements of the kaonic hydrogen and deuterium strong interaction induced shift and width of the 1s levels, with respect to the purely electromagnetic calculated values~\cite{IwasakiPRL1997}.

In 2019, the SIDDHARTA-2 experiment, a major upgrade of SIDDHARTA, has been installed on DA$\Phi$NE aiming to measure, for the first time, the kaonic deuterium 1s level shift and width with a precision similar to that achieved by SIDDHARTA for kaonic hydrogen~\cite{BazziPLB2011a}. The measurement is a great experimental challenge, since the kaonic deuterium X-ray yield is expected to be one order of magnitude smaller than the hydrogen one, and the K$^{-}$ transition lines are expected to be rather broad ($\sim$~1 keV). Therefore, a crucial goal of the new SIDDHARTA-2 detection system is to significantly improve the signal-to-background ratio. To do so, one option is to increase the accepted solid angle by reducing the distance between the deuterium target and the interaction point (IP). The SIDDHARTA-2 setup requires to eliminate the old DA$\Phi$NE luminometer~\cite{BoscoloEPAC2008}. To measure the machine delivered luminosity, SIDDHARTA-2 has been equipped with a new dedicated luminosity detector, fundamental not only for the precise counting of the produced K$^{\pm}$ pairs, but also for the optimization of the DA$\Phi$NE collider beam optics by providing a real-time K/MIPs ratio. 

This paper presents the SIDDHARTA-2 luminosity monitors' working principle, its performances and the first results obtained inside DA$\Phi$NE. 

\section{Detector geometry}

The SIDDHARTA-2 luminosity detector has been designed and constructed in collaboration with the Jagiellonian University in Krak\'ow, Poland. It consists of two parallel detection modules located at the left and right sides of the DA$\Phi$NE beam pipe, in the collision plane, corresponding to the $\phi'$s boost and anti-boost directions~\cite{VignolaCPC1996}, placed at a distance of 7.2~cm from the IP. Each module consists of a 
(80$\times$40$\times$2)~mm$^{3}$ Scionix BC-408 organic scintillator coupled at both ends to fast R4998 Hamamatsu vacuum tube photomultipliers (PMTs) through 6~cm fish-tail plastic lightguides.~The lightguides-scintillator contact surfaces have an angle of 38$^{\circ}$ to fulfill the geometrical constraints of the SIDDHARTA-2 apparatus. 

\begin{figure}[h!]
\centering
	 \includegraphics[width=0.7\textwidth]{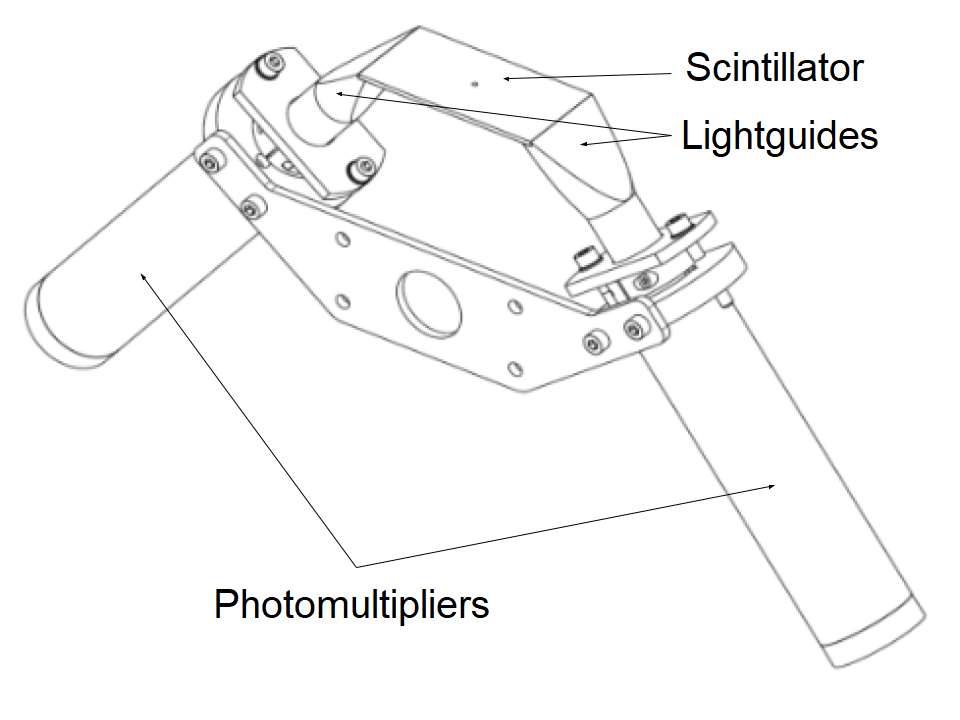}
	\vspace{-0.5cm}
    \caption{Schematic representation of one module of the SIDDHARTA-2 luminosity monitor.}
    \label{fig:Schem_lum}
\end{figure}

\begin{figure}[h!]
\centering
\includegraphics[height=10.0cm,width=12.0cm]{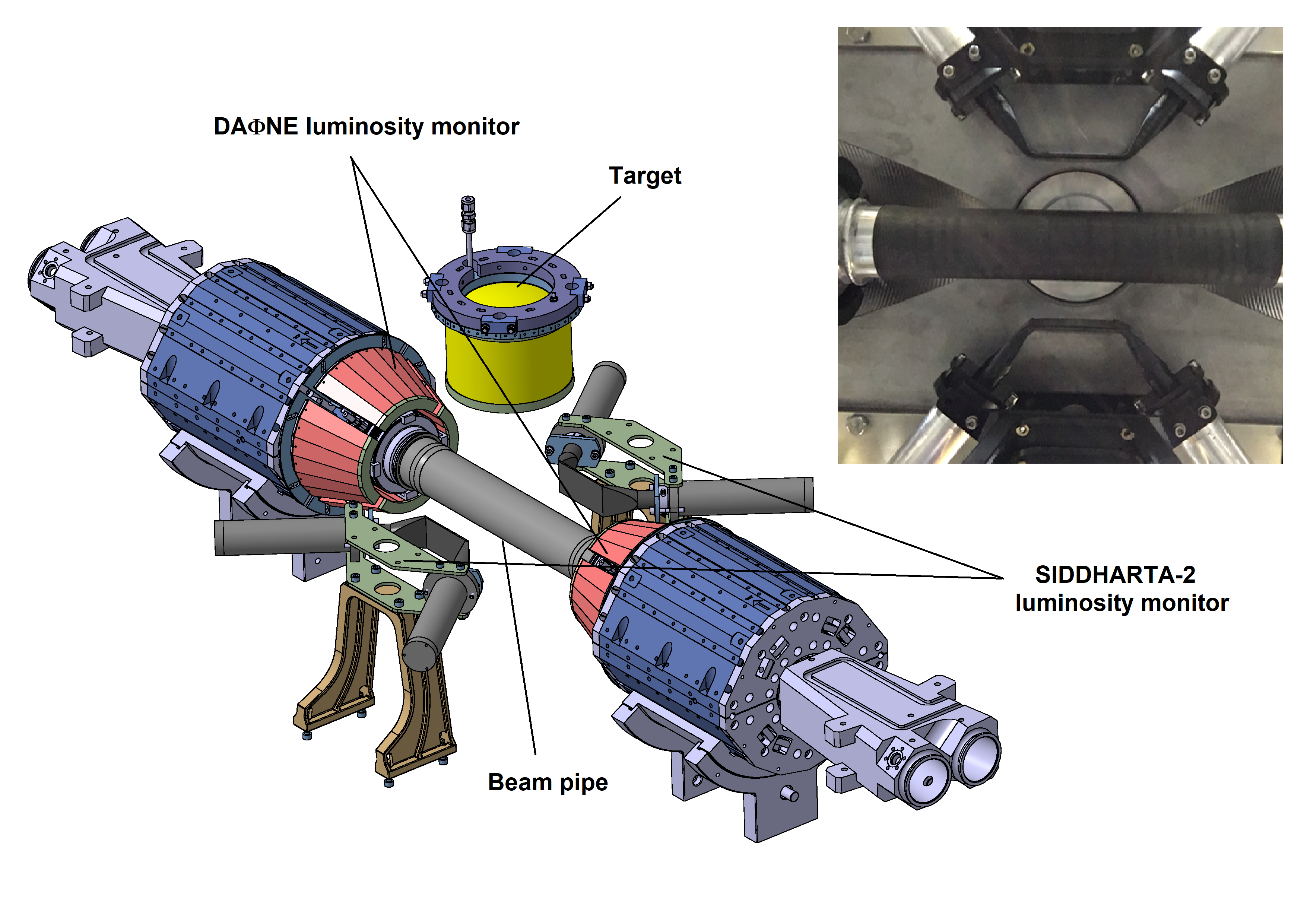}
	\vspace{-1.0cm}
    \caption{Schematic representation of the SIDDHARTA-2 setup with implemented luminosity monitor and the top view picture of the two installed modules (right upper corner).}
    \label{fig:sidd_lum}
\end{figure}

The scintillators and the lightguides are wrapped with reflective and light proof aluminized mylar foils, while the photomultipliers are covered by a tube of aluminum + $\mu$-Metal~(0.1~mm). Each module is equipped with a mechanical support which allows for an easy integration with the experimental setup. The schematic view of the single module and of the SIDDHARTA-2 setup with installed luminosity monitor are presented in Fig.~\ref{fig:Schem_lum} and in Fig.~\ref{fig:sidd_lum}, respectively. 

\section{Laboratory characterization}\label{dt}

Before the integration within the SIDDHARTA-2 experimental setup, the constructed modules underwent laboratory testing. The tests have been performed using a $^{90}\hspace{-0.03cm}\mbox{Sr}$ $\beta$ source to determine the time resolution of each detection module.

The experimental setup consisted of the luminosity detector modules, a LeCroy MSO 44MXs-B oscilloscope and the radioactive source.~The source emitting $\beta$ particles with energies up to $\sim$~2~MeV was held in an aluminum housing with a 2 cm diameter exit window, and was directed at the center of the scintillator.~The PMTs were supplied with several sets of voltages in the 1.9-2.4~kV range in order to investigate the impact on the time resolution; after setting one PMT to a specific value, the other was tuned to the HV value providing the same gain. Gain curves were determined as described in Ref.~\cite{Bednarski_BAMS2014}. 
Data from the two PMTs were acquired by the oscilloscope in a coincidence time window of 25 ns.

To improve the matching with the final SIDDHARTA-2 conditions, the time resolution of each module has been evaluated only for signals with a number of photoelectrons $N_{e}\in(80,120)$, corresponding to the expected deposited energy range of the DA$\Phi$NE kaons and MIPs.

\begin{figure}[h!]
	\centering
    \includegraphics[height=4.5cm,width=6.0cm]{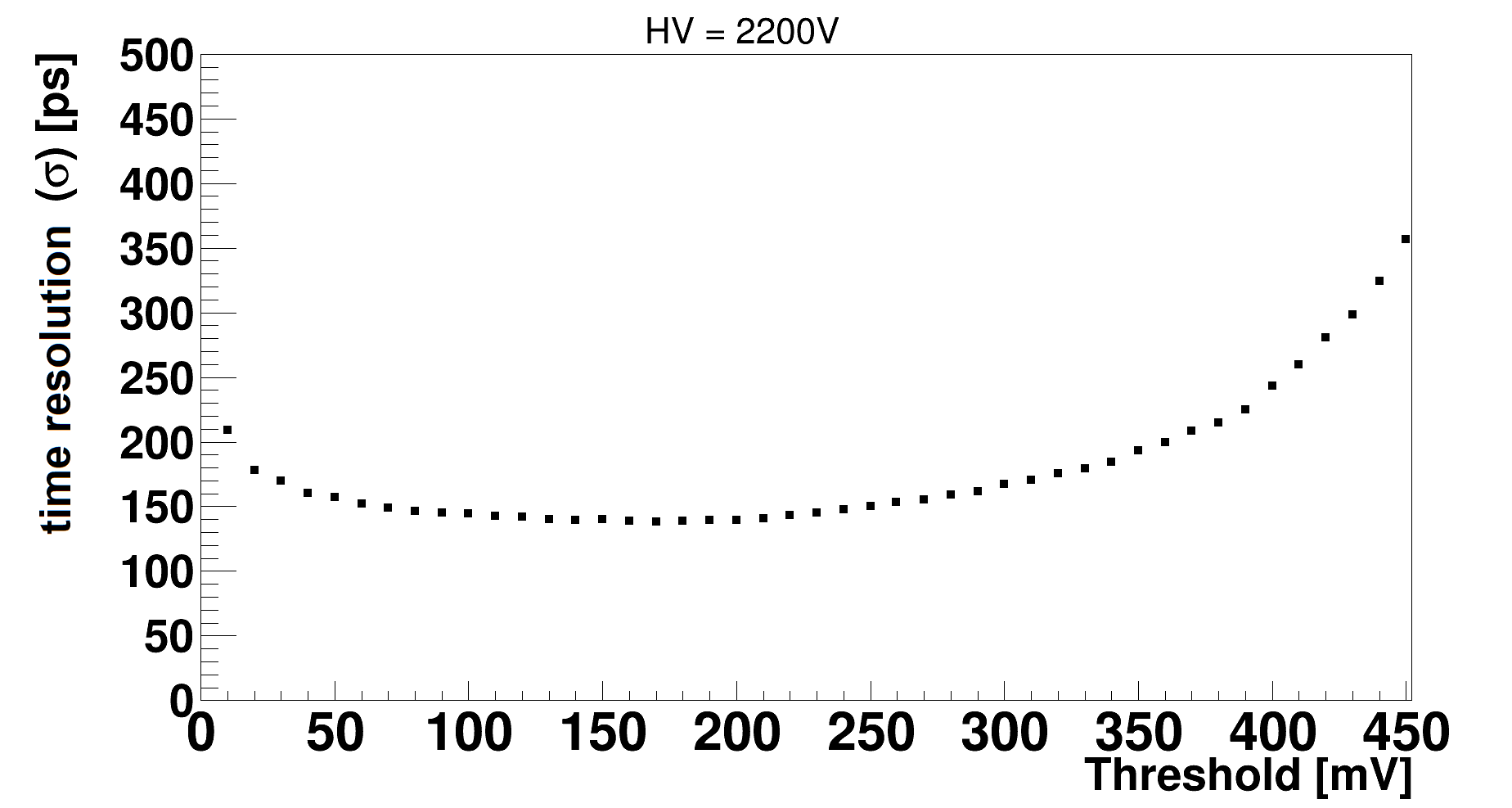}
    \includegraphics[height=4.5cm,width=6.0cm]{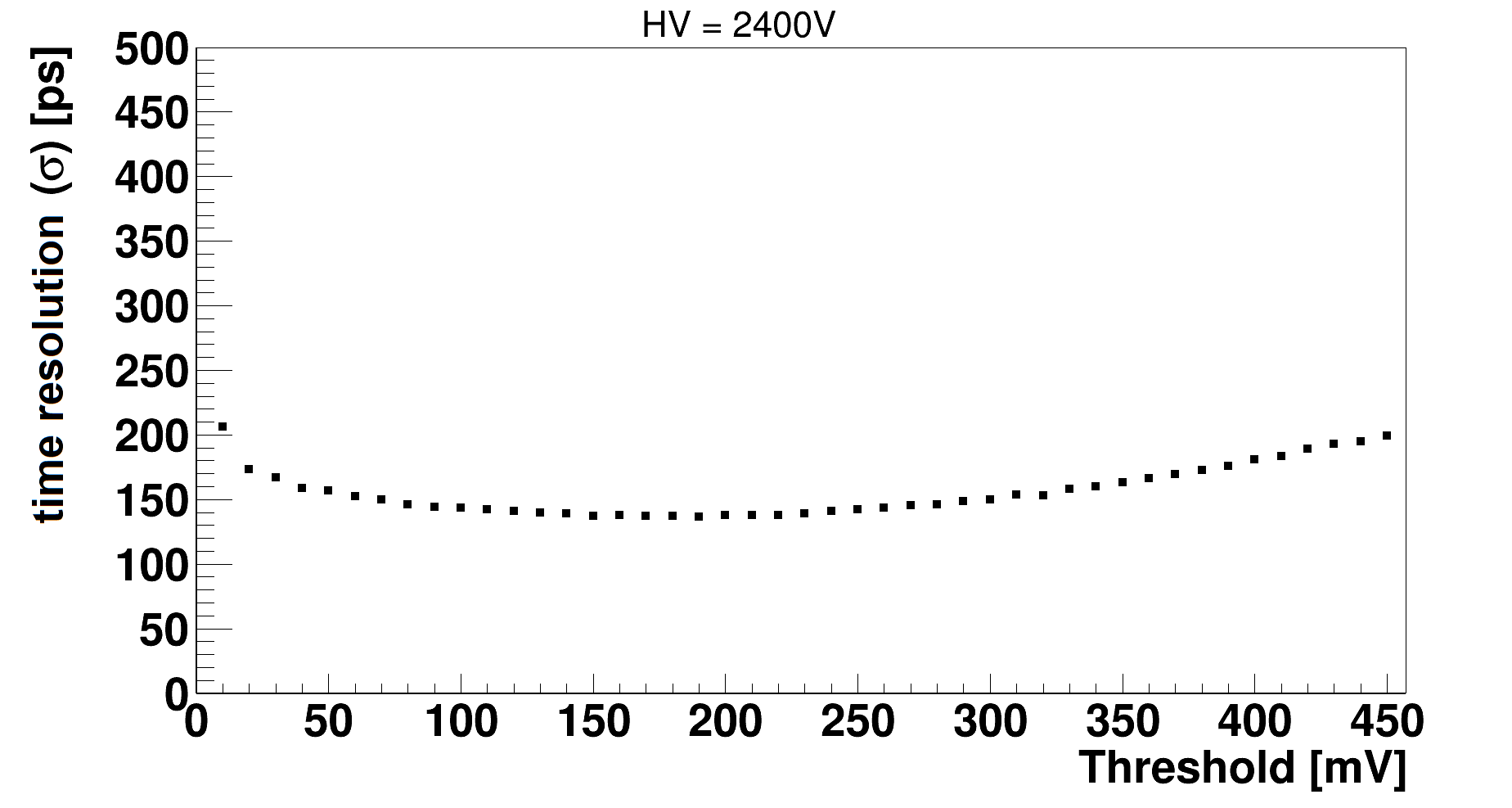}\\
    \vspace{0.5cm}
    \includegraphics[height=4.5cm,width=6.0cm]{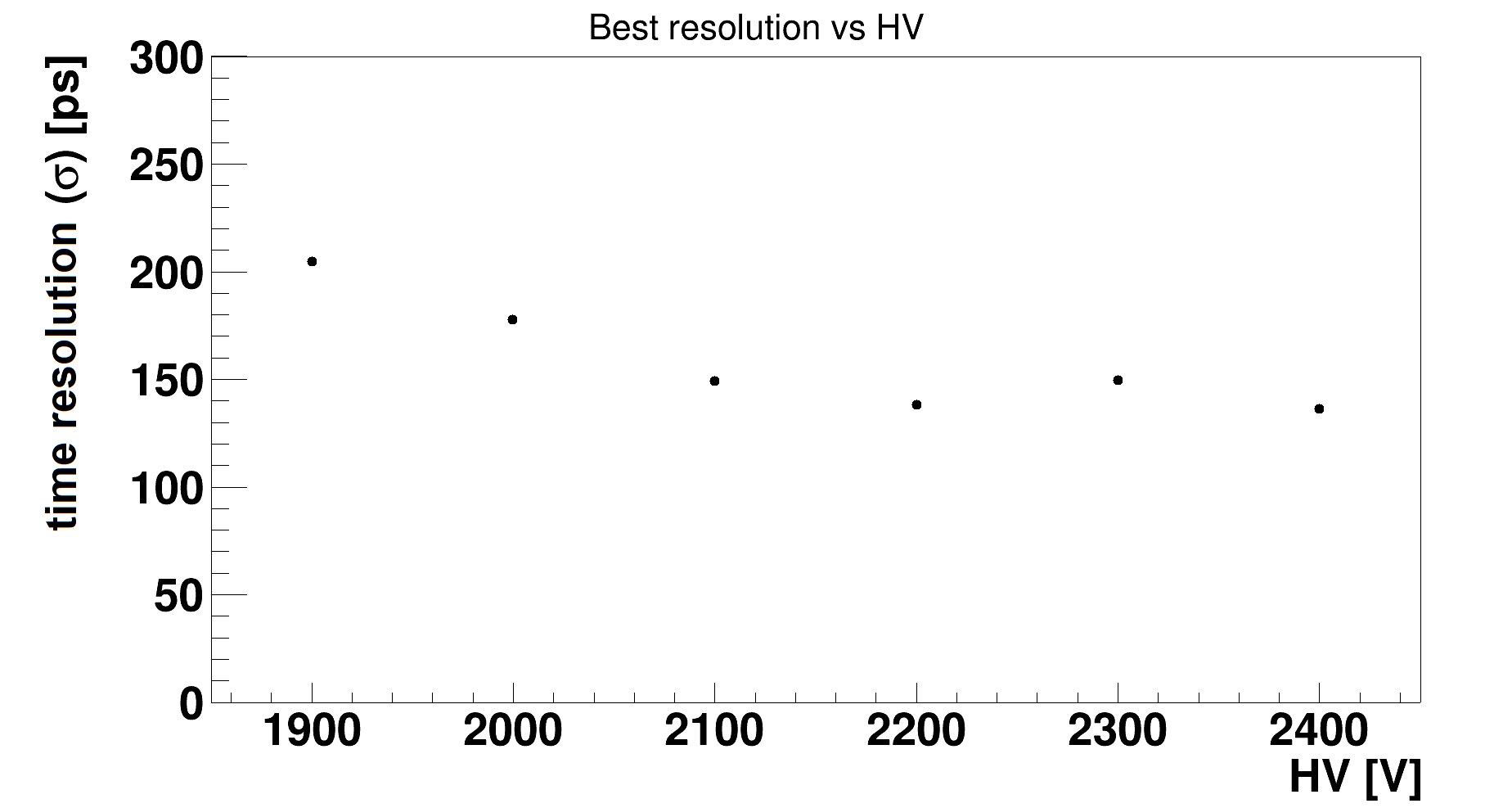}
    \includegraphics[height=4.5cm,width=6.0cm]{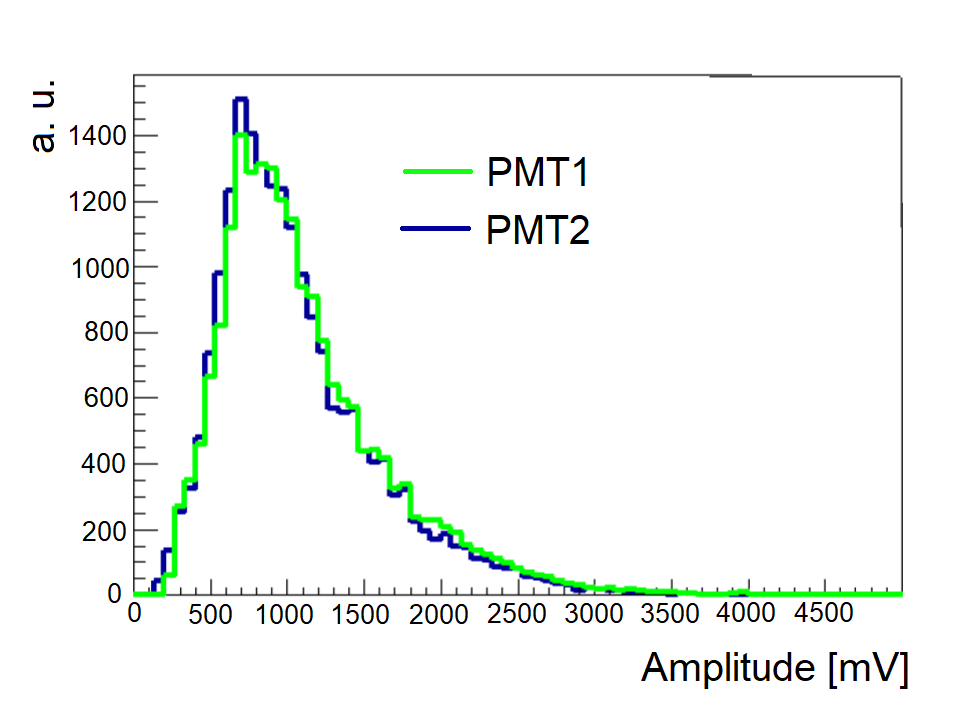}
    \caption{Example of the MT resolution as a function of the threshold value for two sets of HV: 2250V (upper left) and 2394V (upper right) and the best time resolution as a function of HV (lower left). Amplitude of the signals from both PMTs for HV=2400~V (lower right panel).}
    \centering
	\label{ResThr}
\end{figure}

A scan of the Mean Timer (MT) time resolution of each module as a function of the signal threshold has been performed; for each value, the MT distribution was produced, where MT is defined as:
\begin{equation}
t_{MT} = \frac{t_1 + t_2}{2}    
\end{equation} 
with $t_1$ and $t_2$ being the threshold crossing times of each PMTs. The time resolution was retrieved from the $\sigma$ of a gaussian fit of the obtained distributions. As an example, results for two HV values are reported in Fig. \ref{ResThr} (top left and right). For each voltage value, the threshold value corresponding to the best time resolution can be determined. The MT resolution values obtained in this way are shown in the lower pad of Fig.~\ref{ResThr}, as function of the HV values. MT resolution values as low as $\sim$140 ps ($\sigma$) have been obtained.

The best resolution, $\sim$140~ps ($\sigma$) (330~ps (FWHM)), has been obtained for a HV voltage of 2400~V with an average amplitude of the signals from both PMTs of $\sim$800~mV corresponding to a rise time of $\sim$2~ns, a fall time of $\sim$7~ns, and a number of $\sim$100 photoelectrons.

\begin{figure}[!htp]
\centering
\includegraphics[width = 0.7\textwidth]{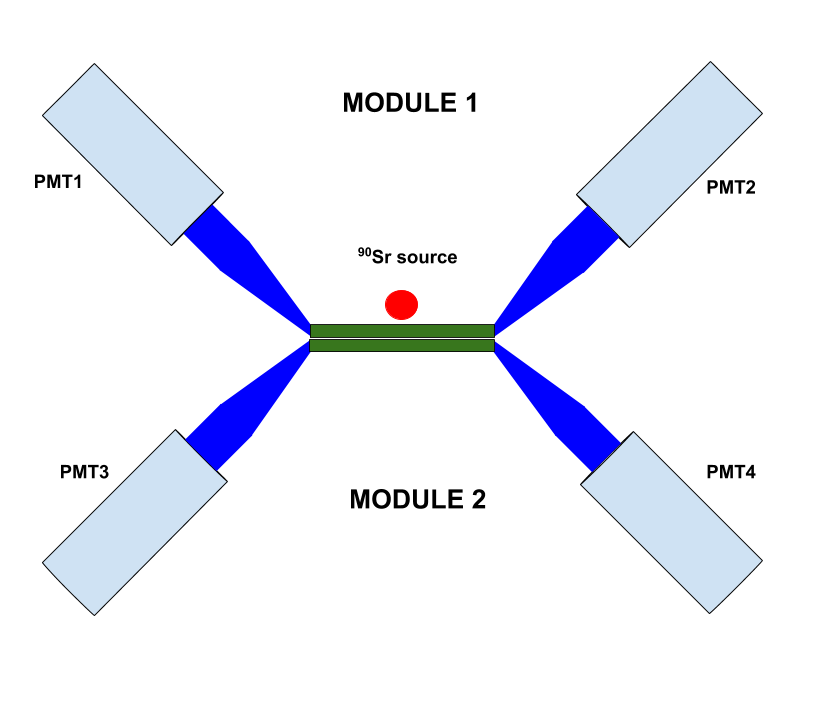}
\vspace{-0.5cm}
\caption{Scheme of the detection setup for the efficiency determination for PMTs in Module~1.}
\label{eff}
\end{figure}

The efficiency of each module has also been evaluated with the $^{90}$Sr source and with the setup shown in Fig. \ref{eff} with the reasonable assumption of a source spot size much smaller than the scintillator surface. The efficiency of Module 1 has been evaluated as the ratio between the counts on Module 1 itself and those on Module 2, since particles hitting the latter are assumed to have passed through the first one too. The efficiency of Module 2 has been evaluated in the same way by inverting the modules. 

The obtained efficiencies vary from about 96\% to 99\% qualifying, together with the MT resolutions, the proposed device as a good luminosity monitor for the SIDDHARTA-2 experiment.

\section{The Data Acquisition System (DAQ)}\label{sec_daq}

The luminosity monitor's main task is to identify the K$^-$ produced in the $\phi$ decay and to reject the MIPs produced during the $e^+e^-$ collisions (mostly pions and electrons) and by $e^+e^-$ lost from the circulating beams. The close distance from the machine IP, necessary to improve the solid angle acceptance, comes with a very high background synchronous with the signal events, as a critical drawback. To enhance the kaon identification, the luminosity monitor has to fulfill four principal requirements:
\begin{enumerate}
    \item Both timing and energy information have to be acquired; the timing information is used for online analysis while the deposited energy can be used for further offline refinement.
    \item A time resolution of $150$~ps$\,(\sigma)$ is needed to ensure kaon identification by time of flight (TOF).
    \item The whole DAQ software chain has to be able to work at trigger rates as high as a few kHz; although the kaon pair production rate during the collision runs on the horizontal plane is expected to be in the range of few Hz, DAQ rates are expected to grow to kHz rate values during DA$\Phi$NE $e^+/e^-$ injection phases.
    \item The DAQ software has to be implemented with an online analysis tool able to provide SIDDHARTA-2 and DA$\Phi$NE, without affecting the DAQ efficiency, with the integrated number of kaons each 15 seconds in order to evaluate the instantaneous luminosity.
\end{enumerate}

\begin{figure}[h!]
\centering
	 \includegraphics[width=10cm]{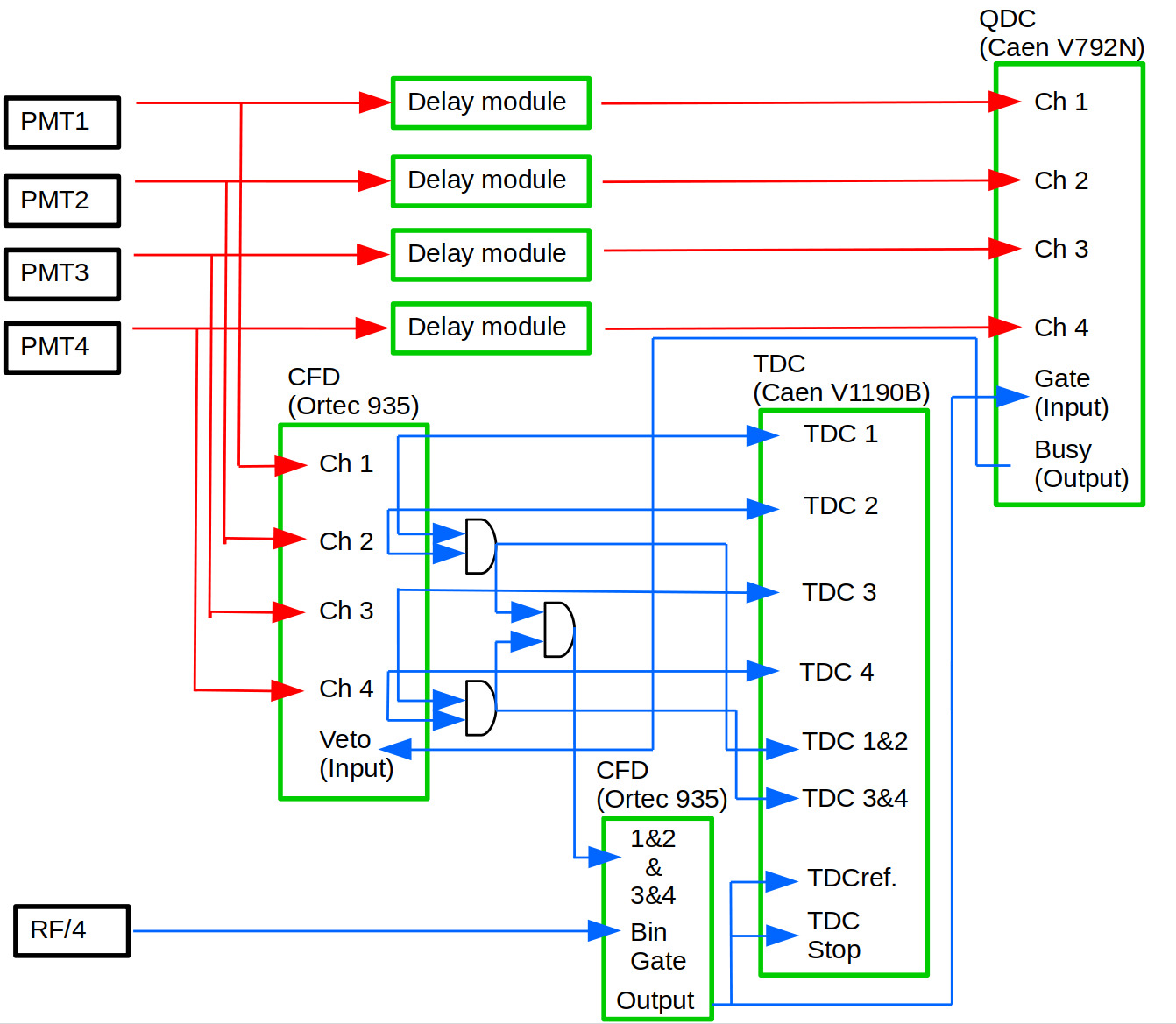}
    \caption{Schematic of the SIDDHARTA-2 luminometer DAQ chain. Red and blue lines represent analog and digital signals, respectively.}
    \label{fig:LumiDAQ_scheme}
\end{figure}

A schematic of the whole DAQ chain is shown in Fig. \ref{fig:LumiDAQ_scheme}

The analog signals from the four PMTs are sent to a passive splitter to double them. The first copy of each signal is then sent to a properly set delay module in order to match the gate window of the CAEN V792N QDC used for the charge integration, while the second copy is used to generate the DAQ trigger logic and to be finally fed to a CAEN V1190B 100 ps resolution TDC module to gather timing information, according to the following logic:

\begin{itemize}
    \item Each of the four PMTs analog signals is sent to a four channel ORTEC 935 Constant Fraction Discriminator, properly tuned to reduce the time walk effects.
    \item The digital outputs are processed by a Coincindence Module (CAEN Mod.~405) (AND logic symbols in Fig.~\ref{fig:LumiDAQ_scheme}) delivering the 1\&2, 3\&4 and 1\&2\&3\&4 coincidence signals.
    \item The 1\&2\&3\&4 digital coincidence is set, using the Bin Gate input of another ORTEC 935CFD module, in coincidence with the RF/4 signal coming from the DA$\Phi$NE collider; this latter is a frequency four times reduced copy of the machine radiofrequency (RF) signal, representing a precise clock of the $e^+e^-$ collision time. The original full RF, having a frequency of 368,7~MHz, couldn't be used due to the CFD module limitations.
    \item The output of this latter coincidence is used both as Gate of the QDC and as common stop for the TDC. The QDC window is 120~ns wide, while the TDC window is 250~ns wide with an offset of -100~ns with respect to the common stop.
    \item The individual four discriminated signals from the PMTs, as well as the 1\&2, 3\&4 and 1\&2\&3\&4 (TDCref. in Fig. \ref{fig:LumiDAQ_scheme}) ones, are acquired by the TDC.
    \item Both the QDC and the TDC readouts are bufferized to ensure the maximum rate capability reducing data reading time; to mantain the charge-time information correlation for each event, the BUSY output of the QDC is sent as a VETO to the first CFD module to inhibit the logic during the QDC digitalization process.
\end{itemize}

\begin{figure}[h!]
\centering
	 \includegraphics[width=9.5cm,height=6.0cm]{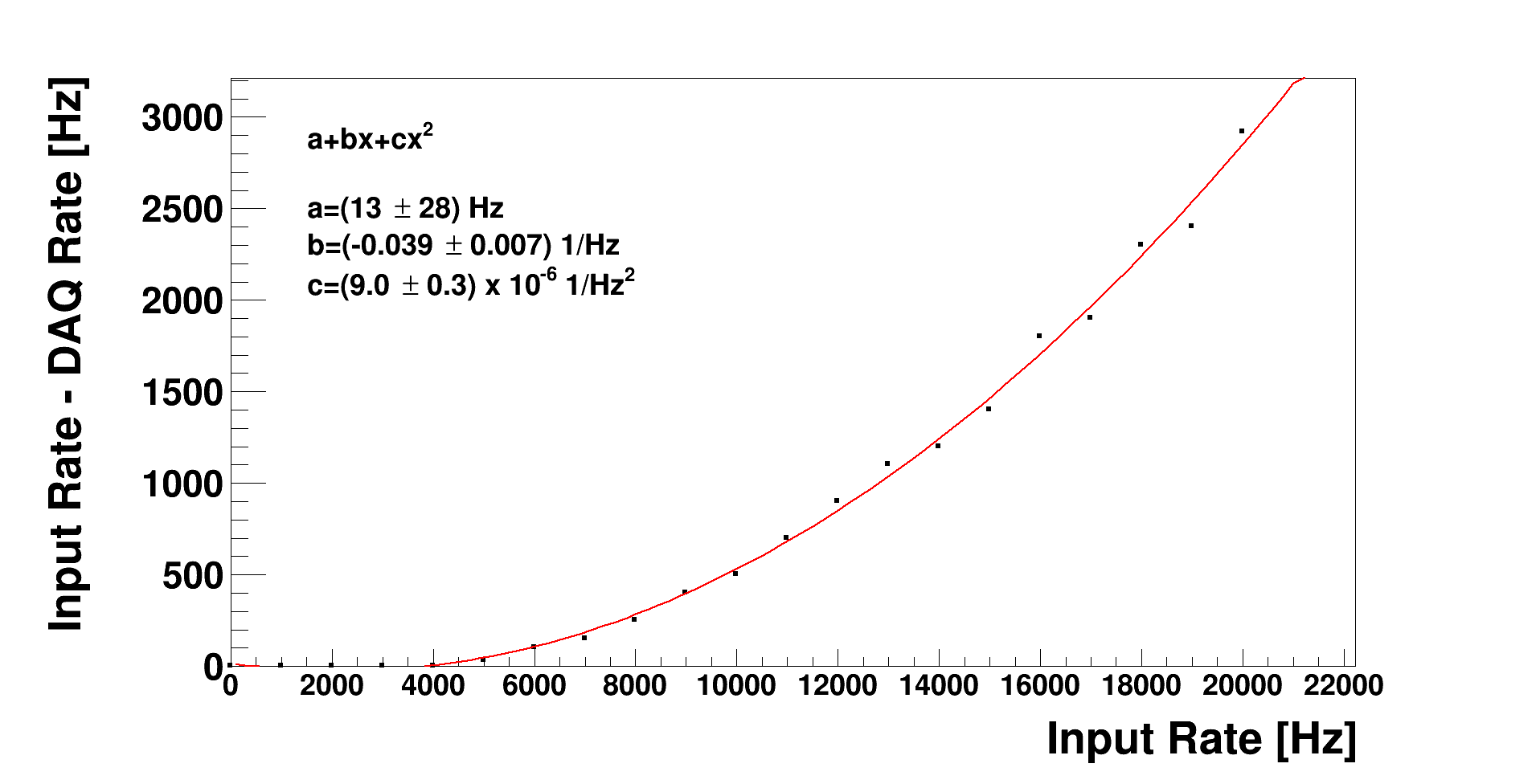}
    \caption{Scan of the the difference between the input and the acquisition rates as a function of the logic pulser input rate. The difference till 4000 input rate is 0. Above the data points are fitted with 2nd degree polynomial function.}
    \label{fig:DAQ-rates}
\end{figure}

The rate capability of the DAQ chain has been tested with a logic pulser sending four identical signals, having the same shape (amplitude, rise time and duration) of the typical kaon/MIPs signals, to the four splitters leaving the rest of the logic unchanged. During this test the software part used to analyze the data buffer each 15 seconds period of data taking to deliver online K and MIPs counts was also running to test its influence on the DAQ chain. A scan of the acquisition rate as a function of the logic pulser input rate has been performed and the results are shown in Fig.~\ref{fig:DAQ-rates}, where the difference between the input and the acquisition rates is plot versus the input one.

\begin{figure}[h!]
\centering
	 \includegraphics[height=4.5cm,width=6.0cm]{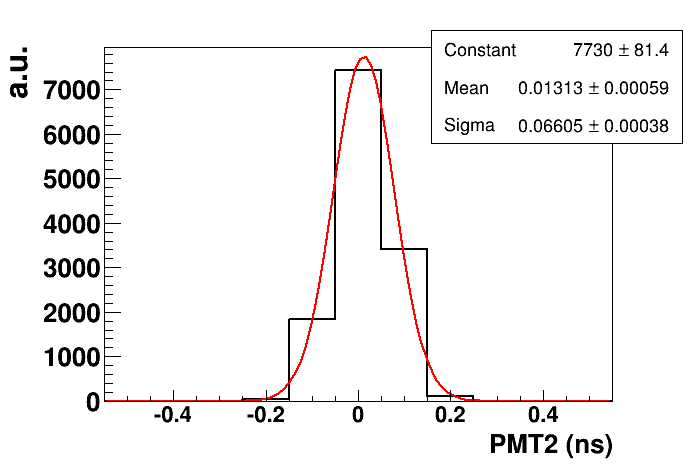} \\
	 \includegraphics[height=4.5cm,width=6.0cm]{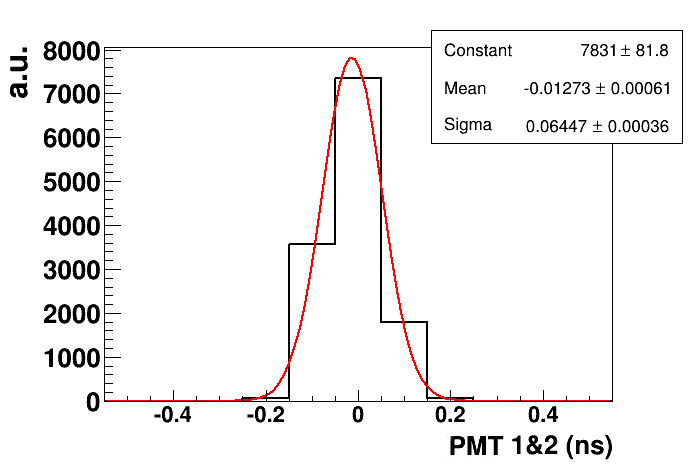}
	 \includegraphics[height=4.5cm,width=6.0cm]{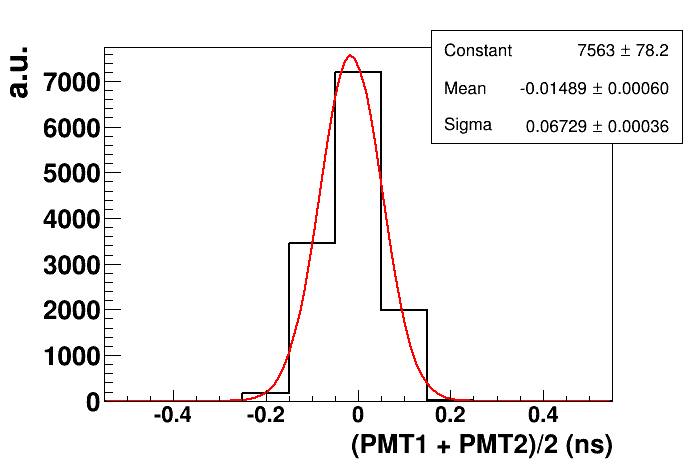}
    \caption{DAQ chain induced tests: time distribution of PMT2 (up), of the PMT1\&PMT2 (bottom left) and of the offline reconstructed Mean Timer (PMT1+PMT2)/2 (bottom right).}
    \label{DAQ-pulser}
\end{figure}

The obtained results ensure that no events are lost for trigger rates up to 4~kHz. Using the same logic, since the four signals from the pulser are produced simultaneously, the DAQ chain induced jitter could be evaluated looking at the time distribution of each of the acquired TDC channels; the results are presented in Fig. \ref{DAQ-pulser} where, as an example, the time distribution of the PMT2 (up), of the PMT1\&PMT2 (bottom left) as well as the one of the offline reconstructed Mean Timer (PMT1+PMT2)/2 (bottom right) are reported. The obtained time resolutions of $\sigma \simeq 60\,$ps ensure a negligible impact on the DAQ chain.  

\section{First results obtained inside DA$\Phi$NE}

\begin{figure}[h!]
\centering
\includegraphics[height=8.0cm,width=8.5cm]{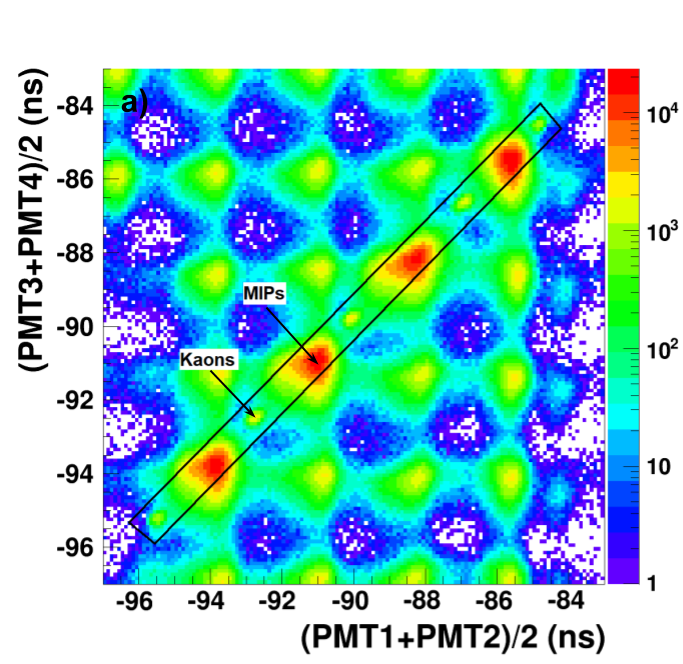}
\includegraphics[height=8.0cm,width=8.5cm]{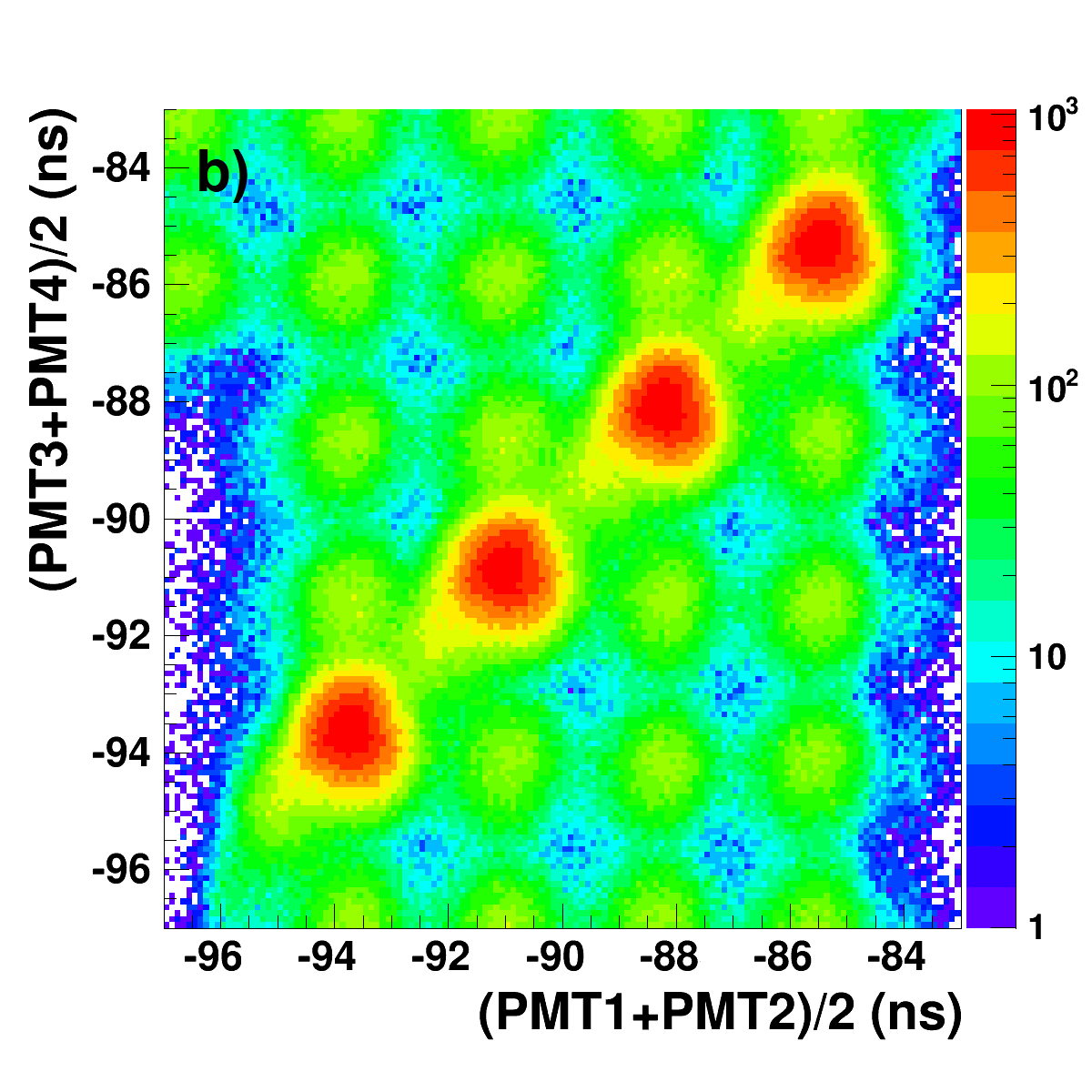}
\caption{2D plot representing the average TDCs for the signals in coincidence registered on the boost ((PMT1+PMT2)/2) and anti-boost ((PMT3+PMT4)/2) sides of IP for collision (a) and no-collision modes (b). The difference of the background levels at the kaons location results from the different DA$\Phi$NE conditions during the run with and without collisions.}
\label{time_spec1}
\end{figure}

\begin{figure}[h!]
\centering
\includegraphics[height=7.5cm,width=12.0cm]{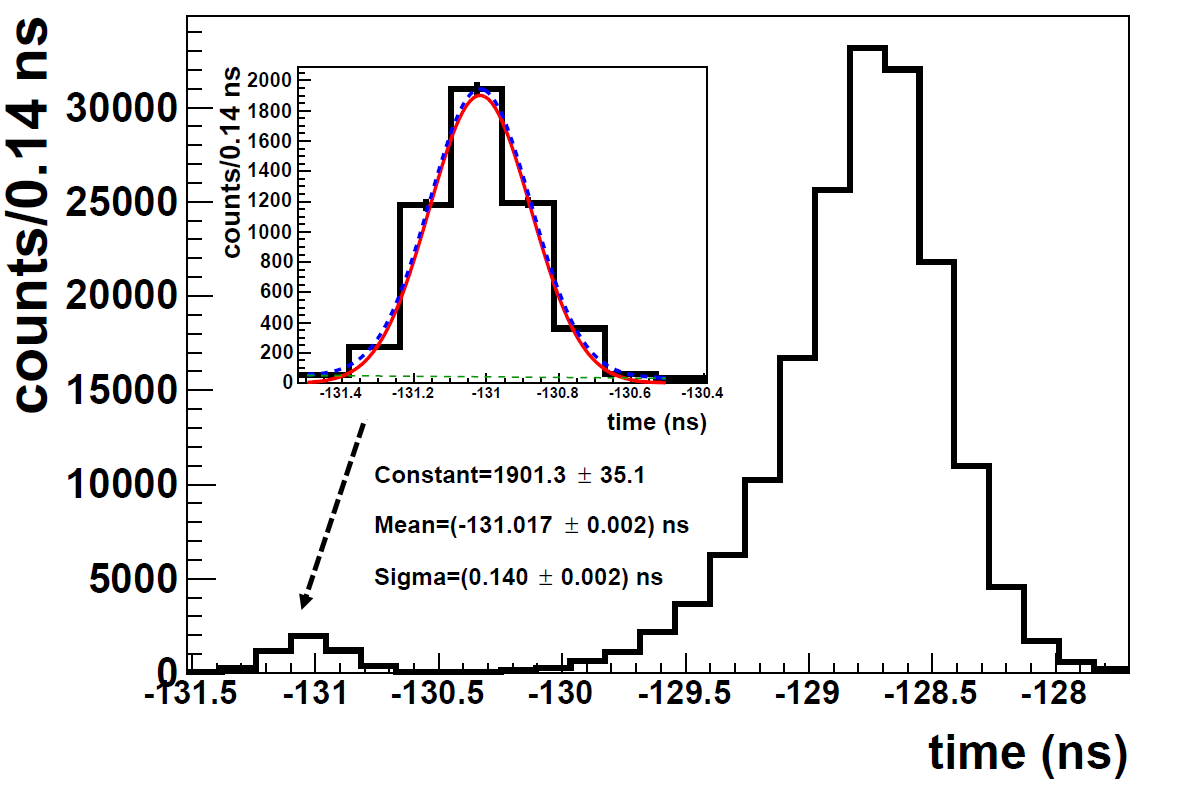}
\caption{Projection on the diagonal time coordinate of the 2D plot. The inner pad shows the fit to kaon peak (blue dashed line). The signal and background components are marked with the solid red and dashed green lines, respectively.}
\label{time_spec2}
\end{figure}

We present here the first results obtained with the SIDDHARTA-2 luminosity detector during the ”DA$\Phi$NE commissioning phase”. 
The data correspond to two hours data taking (00:36 - 02:45 16/02/2020); the measurements were performed with the beam optics set both in $e^{+}$ $e^{-}$ collision and non-collision mode to better disentangle between the kaon signal (from $\phi$'s decay) and the background (mainly MIPs due to beam losses). The distance between the luminometer and the interaction point (7.2~cm) was optimized, based on K$^-$ and MIPs TOF, to provide the clearest separation between the two components. Fig.~\ref{time_spec1}(a) and (b) show the scatter plot for the Mean Time measurements of the two detector modules for collision and no-collision mode, respectively. The two modules were placed on the boost ((PMT1+PMT2)/2) and anti-boost ((PMT3+PMT4)/2) sides of the IP. On the diagonal in Fig.~\ref{time_spec1}(a), four double stuctures can be identified, due to the RF/4 trigger signal, each consisting in two clearly distinguishable bumps corresponding to kaons and MIPs. Similar bumps are also visible out of the diagonal, originated by particles belonging to different RF bunches but still matching the coincidence time window; only MIPs contribute to these regions because of the negligible probability to find two kaons in coincidence from subsequent RF bunches. Comparing the scatterplots obtained with and without collisions, the kaons region can be immediately identified. The number of kaons ($N_{K}$) produced in the beam collisions, crucial for the luminosity determination, can be extracted based on the projection on the diagonal, by fitting the kaon peaks with a gaussian distribution and the background with a second-degree polynomial. As an example, the fit for a single structure is shown in Fig.~\ref{time_spec2}.

The number of kaons is determined taking into account both the geometrical efficiency, estimated by Monte Carlo simulations to be 5.7\%, and the overall detection efficiency of the luminosity monitor (90\%) calculated from those obtained for the single modules (see Sec. \ref{dt}). The luminosity can be then estimated according to the expression:

\begin{equation}
L=\frac{N_{K}}{t_{DAQ} \cdot \sigma_{e^+e^-\rightarrow\phi} \cdot BR_{\phi\rightarrow K^+K^-}}   
\end{equation}

\noindent where $t_{DAQ}$ denotes the effective time of data taking (excluding injection), $\sigma_{e^+e^-\rightarrow\phi}$ is the cross section for $e^+e^-\rightarrow\phi$ process measured by SND collaboration~\cite{Achasov2001} and $BR_{\phi\rightarrow K^+K^-}$ is the Branching Ratio for $\phi$ meson decay to K$^+$K$^-$ pair equal to 48.9$\pm0.5\%$~\cite{pdg}. 
The mean luminosity value obtained for the mentioned data sample is (3.26$\pm$0.05(stat.)$^{+0.17}_{-0.15}$(syst.))$\cdot$10$^{31}cm^{-2}s^{-1}$. The systematic errors originate from the uncertainty of the $\sigma_{e^+e^-\rightarrow\phi}$ as well as from the MC calculated impact of a maximum error of $\pm$2mm in the luminosity monitor distance from the IP with respect to the nominal one. The total, positive and negative systematic errors are obtained summing the contributions in quadrature.

\begin{figure}[h!]
\centering
\includegraphics[height=6.0cm,width=10.5cm]{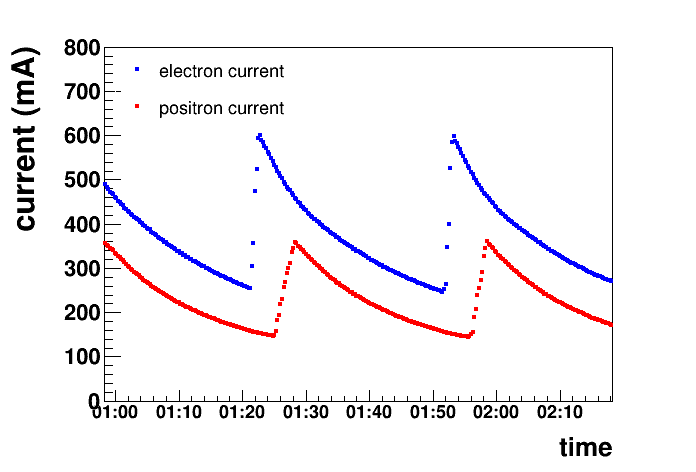}\\
\includegraphics[height=6.0cm,width=10.5cm]{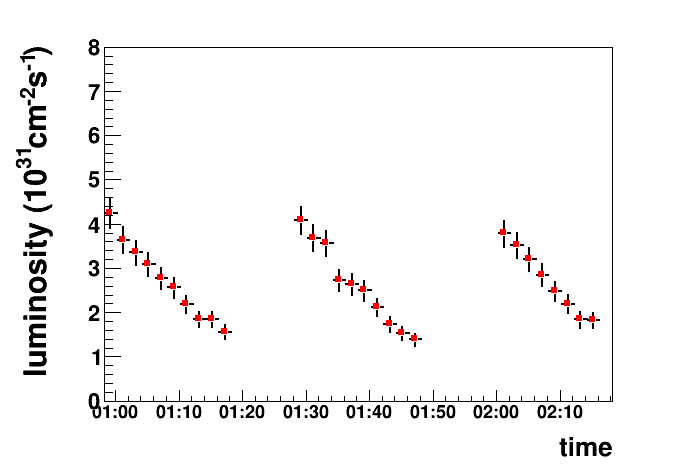}
\caption{(upper) DA$\Phi$NE currents: electron (blue) and positron (red); (lower) measured luminosity - each point corresponds to 2 min of data taking.}
\label{lum_dafne}
\end{figure}

The aim of the luminosity monitor system is to provide value of the luminosity of each DA$\Phi$NE beam cycle during the experimental run. As an example the luminosities for three beam cycles in the considered data sample, are shown in the lower part of Fig.~\ref{lum_dafne}. The luminosity was evaluated using 2~min time intervals and excluding the beam injection periods where very high background condition of the DA$\Phi$NE machine results in hundreds kHz rate on the luminometer saturating the DAQ system. Within each beam cycle the luminosity decreases following the lifetime of the electron and positron beam currents (the upper part of Fig.~\ref{lum_dafne}). Presently DA$\Phi$NE is in optimization phase, mostly dedicated to increase the luminosity, which will be followed by an optimization aiming to reduce of the background. During the SIDDHARTA-2 experiment, due to the foreseen lower background conditions, the luminosity evaluation will be possible also during injections. Online values of the luminosity will be delivered to the DA$\Phi$NE team for the beams collisions monitoring and optimization.

\section{Conclusions}

We presented the SIDDHARTA-2 luminosity detector system, characterized by high efficiency and excellent time resolution, with a dedicated DAQ designed to fulfill all the requirements of SIDDHARTA-2 kaonic deuterium experiment at DA$\Phi$NE. The tests performed during the DA$\Phi$NE commissioning phase show that the luminosity detector can be used for the measurement of luminosities, having measured values in the order of 10$^{31}$cm$^{-2}$s$^{-1}$, about one order of magnitude lower than expected during SIDDHARTA-2 dedicated run. The measured values are perfectly in line with what expected in the phase of commissioning and with past DA$\Phi$NE performance in equivalent beam conditions~\cite{dear}. We have then tested and validated a main tool, both for SIDDHARTA-2 and DA$\Phi$NE, in the preparation of the kaonic deuterium challenging measurement.

\section*{Acknowledgements}
We thank C. Capoccia from LNF-INFN and H. Schneider, L. Stohwasser, and D. Pristauz-Telsnigg from Stefan-Meyer-Institut for their fundamental contribution in designing and building the SIDDHARTA-2 setup.~We thank as well the DA$\Phi$NE staff for the excellent working conditions and permanent support. Part of this work was supported by the Austrian Science Fund (FWF): [P24756-N20 and P33037-N]; the Croatian Science Foundation under the project IP-2018-01-8570; EU STRONG-2020 project (grant agreement No.~824093) and the Polish Ministry of Science and Higher Education grant No. 7150/E-338/M/2018.

\bibliographystyle{unsrt}
\end{document}